# Unmasking the Origin of Kinks in the Photoemission Spectra of Cuprate Superconductors

Zhenglu Li, Meng Wu, Yang-Hao Chan, and Steven G. Louie*

*Department of Physics, University of California at Berkeley, Berkeley, CA 94720, USA and Materials Sciences Division, Lawrence Berkeley National Laboratory, Berkeley, CA 94720, USA.*

*Email: sglouie@berkeley.edu

## Abstract

The origin of a ubiquitous bosonic coupling feature in the photoemission spectra of high-$T_c$ cuprates, an energy-momentum dispersion 'kink' observed at ~70 meV binding energy, remains a two-decade-old mystery. Understanding this phenomenon requires an accurate description of the coupling between the electron and some collective modes. We report here *ab initio* calculations based on *GW* perturbation theory and show that correlation-enhanced electron-phonon interaction in cuprates gives rise to the strong kinks, which not only explains quantitatively the observations but provides new understanding of experiments. Our results reveal it is the electron density-of-states being the predominant factor in determining the doping-dependence of the kink size, manifesting the multi-band nature of the cuprates, as opposed to the prevalent belief of it being a measure of the mode-coupling strength.

## Main text

Rich exotic phases in copper oxide superconductors (the cuprates) arise from correlated electrons and their interactions with other elementary excitations [1]. Over the past two decades, intensive angle-resolved photoemission spectroscopy (ARPES) experiments have established a major bosonic coupling fingerprint in the cuprates – namely, along the *d*-wave superconducting gap nodal direction in the reciprocal space, the quasiparticle dispersion relation kinks at a binding energy around 70–80 meV [2–4]. A phonon mechanism was initially conjectured [3] because the kink's energy matches that of the oxygen breathing phonon energy [5–9] and later supported by multiple-mode features [10] and the isotope effect [11,12]. On the other hand, spin fluctuations were also proposed as an alternative mechanism [4,13], as well as a boson-free mechanism that had been suggested [14]. The debate on the physical origin of the kink is yet to be settled because there are various collective excitations or bosonic modes in the cuprates, any of which may lead to a similar phenomenon with properly fitted model parameters [9,13,14] to experimental data. To provide unbiased conclusions, predictive first-principles approaches (with no adjustable parameters) become necessary.

Knowing the origin of the photoemission kink would provide new insights to interactions in this important class of materials and may lead to better understanding of other phenomena including superconductivity. A major impediment has been a lack of truly *ab initio* calculations that can quantitatively explain the kink feature. In particular, within the mechanism of electron-phonon (*e*-ph) coupling induced kink, previous calculations [5,6] – based on *ab initio* density-functional theory (DFT) [15,16,17] – yielded a kink that is nearly a factor of 3 too small in magnitude as compared with experiments. Meanwhile, there are yet to be any *ab initio* studies capable of explaining the kink quantitatively in a parameter-free fashion with other proposed mechanisms. It is however important to note that standard static DFT fundamentally does not address the excited states of materials [18,19]; its naive application to the coupling between the excited quasiparticles and other elementary excitations can be severely misrepresented in some materials [17,20]. Thus, even within the *e*-ph mechanism, it remains unclear whether the observed discrepancy is due to the inadequacy of the standard DFT formulation of the *e*-ph coupling (i.e., treating the electrons as the fictitious Kohn-Sham noninteracting electrons) or indeed phonons are not involved. Here we address this question using a newly developed method – *GW* perturbation theory (*GW*PT) [20] – that has successfully included many-electron correlation effects in computing the *e*-ph interaction from first principles. The *e*-ph coupling is properly computed as the interaction of a true quasiparticle (including nonlocal self-energy effects) with the phonons, within a linear-response and interacting-Green's-function formalism in the *GW* approximation [18–20].

We investigate the *e*-ph interaction and photoemission kinks in the prototypical single-copper-oxygen-layer cuprate $La_{2-x}Sr_xCuO_4$ (LSCO) in the optimally doped ($x = 0.15$) and overdoped ($x = 0.30$) hole concentrations. We do not address the underdoped cuprates in this work because they possess very strong local correlations [1] for which the application of quasiparticle-based *GW*PT formalism may not be properly justified. We find that, correlation effects included in *GW*PT significantly enhances the phonon-induced part of the quasi-electron self-energy by a factor of 2–3 as compared with results from DFT *e*-ph coupling [5,6] as obtained from density functional perturbation theory (DFPT) [16,17]. The calculated kink magnitude, as well as its doping and temperature dependence, quantitatively agree with a broad range of experiments [3,21–26], showing that the nodal photoemission kink in cuprates is mainly caused by *e*-ph coupling. The excellent agreement with experiments from first principles enables us to proceed with confident to dissect the hidden interplays among the *e*-ph interaction, electronic structure, and photoemission kink of the cuprates, which can often be obscured in the experimental data and subjected to speculation. Indeed, from optimally doped to overdoped regime, our results show a reduced kink size in agreement with experiments. However, contrary to the common interpretation [3,4,21–26] attributing this doping dependence of the kink size to a weakening electron-boson coupling strength, our analysis clearly shows that this is in fact due to a multi-band density-of-states (DOS) effect previously not recognized.

An excited electron or hole in a crystal interacts with other electrons and elementary excitations including phonons, acquiring a shift in its excitation energy and a finite lifetime. These many-body interaction effects may be formally cast into a complex operator called the self-energy $\Sigma$ associated with a quasiparticle, and can be measured in ARPES experiments. The *GW* method [18,19,27] expands the electron-electron part of the self-energy to first order in the Green's function ($G$) and in the screened Coulomb interaction ($W$), i.e. $\Sigma^{e-e}(\mathbf{r},\mathbf{r}';\varepsilon) = iGW$, capturing the non-local $(\mathbf{r},\mathbf{r}')$ and frequency ($\varepsilon$) dependence of the interaction. The *e*-ph coupling matrix element describes the amplitude of an electron being coupled from one quasiparticle state to another, due to changes in the potentials (both crystal potential and self-energy) induced by phonons seen by the electron, which can be efficiently evaluated via linear-response theory [16,17,20]. The *GW*PT approach [20] directly computes the changes in the electronic self-energy due to perturbing phonons (with phonon wavevector $\mathbf{q}$ and branch index $\nu$), $\partial_{\mathbf{q}\nu}\Sigma^{e-e}(\mathbf{r},\mathbf{r}';\varepsilon)$, in constructing the *e*-ph matrix elements – a core ingredient of microscopic *e*-ph theories including electron correlation effects. (See theoretical and computational details, along with a *GW*PT validation set of five conventional metals and one oxide metal, in the Supplemental Material [28].)

Having accurate *ab initio* *e*-ph interactions allows us to compute the *phonon-induced* part of the electron's self-energy $\Sigma^{e-\mathrm{ph}}$, which reshapes the electron dispersion relation and provides notable phonon-induced signatures if any (such as kinks in ARPES). For comparison, we calculate the *e*-ph properties of LSCO using both *GW*PT- and DFPT-calculated *e*-ph matrix elements, together with a Wannier interpolation technique [59]. Figure 1 shows the nodal $\Sigma^{e-\mathrm{ph}}_{n\mathbf{k}}(E_{n\mathbf{k}})$ (with electron wavevector $\mathbf{k}$ and band index $n$) computed within the Fan-Migdal approximation [17,60–62] at low temperature ($T = 20$ K). In Figs. 1(a) and 1(b), $\mathrm{Re}\Sigma^{e-\mathrm{ph}}_{n\mathbf{k}}$ shows a dominant peak at 75 meV and 76 meV binding energy for $x = 0.15$ and $x = 0.30$, respectively, whose origin is traced to come from the Cu-O in-plane half- and full-breathing modes as found previously [5]. The calculated peak positions nicely agree with the experimentally measured kink positions [3,21] at around 70–80 meV binding energies. The shoulder structure near 40 meV is due to the oxygen buckling and stretching modes [5]. Beyond the maximum phonon frequency $\omega^{\mathrm{max}}_{\mathrm{ph}} = 87$ meV, $\mathrm{Re}\Sigma_{n\mathbf{k}}^{e-\mathrm{ph}}$ behaves as a featureless flat tail up to 200 meV binding energies without any peaks. Figures 1(c) and 1(d) show a roughly double-step structure in the magnitude of $\mathrm{Im}\Sigma_{n\mathbf{k}}^{e-\mathrm{p}}$ (with step positions corresponding to the abovementioned 40 meV and 75 meV phonon modes) and then slowly decays beyond $\omega_{\mathrm{ph}}^{\mathrm{max}}$ up to 200 meV binding energy [28]. Importantly, for both $x$ = 0.15 and $x$ = 0.30, the peak magnitude of $\mathrm{Re}\Sigma_{n\mathbf{k}}^{e-\mathrm{ph}}$ and $\mathrm{Im}\Sigma_{n\mathbf{k}}^{e-\mathrm{ph}}$ within the phonon frequency range is around 25–35 meV from *GW*PT, but is only 10–15 meV from DFPT, representing a correlation enhancement in the *e*-ph induced self-energy by a factor of about 2–3, which is directly reflected in the kink magnitude as discussed below.

We now explore in detail the low-energy (<100 meV) region in the nodal dispersion relation where phonons produce sharp features, although spectroscopic features exist at higher binding energies [63,64] as well. Experimental extraction of the exact size and position of the kink in the measured quasiparticle dispersion relation depends on the choice of a reference bare band [65]. The common practice is to assume the bare band (i.e., without the effects of bosonic mode coupling of interest) as a straight line that connects the quasi-hole state at Fermi energy $E_F$ (set to zero throughout this paper) to another state at an arbitrarily chosen lower energy $E_1$, which had often taken a range from at –0.1 to –0.3 eV in the literature [3,4,10–12,21–26,63,64,66]. In this practice, the deviation at each $\mathbf{k}$ point of the measured dispersion relation from the assumed reference line is usually mis-labeled as “Re$\Sigma$”, although it is in fact *not* the real part of the self-energy. Here, we shall follow this experimental practice to extract the size and location of the kinks to make direct comparisons between theoretical and experimental data. To emphasize the low-energy bosonic-coupling contributions, we choose $E_1$ = –0.12 eV (similar to many experiments) such that the extracted kink covers all the rapidly varying features of $\mathrm{Re}\Sigma_{n\mathbf{k}}^{e-\mathrm{ph}}$ while staying away from the high-binding-energy features where coupling to other higher-energy collective excitations may play a role [63–65]. We emphasize that the extracted photoemission kink should be distinguished from the true $\mathrm{Re}\Sigma_{n\mathbf{k}}^{e-\mathrm{ph}}$ that are shown in Fig. 1, and these two quantities are related via Eq. (S7) in the Supplemental Material [28].

Figures 2(a) and 2(c) show the calculated electron spectral functions $A_{n\mathbf{k}}(\omega)$, where the quasiparticle self-energy features (the kink and spectral widths) are clearly visible. We plot the momentum distribution curves (MDCs) at regular energy steps [Figs. 2(b) and 2(d)] and fit each to a Lorentzian function to obtain the MDC-derived energy-*vs.*-wavevector dispersion relation and the linewidth. This data processing procedure is well-defined and is widely adopted in analyzing experimental ARPES spectra [67]. After aligning the reference line for *GW*PT, DFPT, and the experimental dispersion relations to eliminate the unknown bare-band information, a direct comparison of the kink features is made between theory and experiment [3] in Figs. 2(e) and 2(f). We find remarkably good agreement between *GW*PT and experiments [3,21] whereas, as found in previous study [5,6], DFPT dramatically underestimates the kink size. Furthermore, the lifetime information embedded in MDC linewidths shows reasonably good agreement between *GW*PT and experiment [22] [Figs. 2(g) and 2(h)], although a sizable uncertainty depending on the exact fitting procedure to experimental data remains [28]. The reason that the *e*-ph interaction from *GW*PT greatly improves the agreement between theory and experiment is because of its enhanced value from inclusion of electron correlation effects [20].

In Fig. 3, we directly compare the extracted kink features with multiple sets of experimental data [3,21,23–26] on LSCO and $Bi_2Sr_2CuO_{6+\delta}$ (Bi2201) (another single-layer cuprate). Quantitative agreement between the experimental kinks and theoretical *GW*PT kinks is clearly seen [Figs. 3(a) and 3(b)], giving a maximum deviation from the reference line of ~20 meV for optimally doped and ~15 meV for overdoped LSCO and Bi2201, mirroring closely those observed in experiments. We further use the area under the kink-amplitude-*vs.*-energy curve [Figs. 3(a) and 3(b)] at different temperatures for a measure of the temperature dependence of the kink strength, where good agreement between *GW*PT and experiments is also found as seen in Figs. 3(c) and 3(d). The kink strength monotonically reduces as the temperature rises. This is because at higher temperature, both phonon absorption and emission processes become more activated, providing increasingly more scattering channels that would broaden the phonon-induced self-energy peaks (Fig. S1), leading to a reduced apparent kink size. The quantitative agreement on the temperature dependence of the apparent kink strength between results from *GW*PT and various experiments consolidates the phonon mechanism being the explanation for the 70–80 meV nodal kink in the cuprates (at least as the dominant origin – it is possible other mechanisms [9,13,14] may also contribute in some way). At temperatures below superconducting $T_c$ (~30–40 K for LSCO and Bi2201 at optimal doping), the existence of a non-zero superconducting gap shifts the quasiparticle energies and may have an additional effect (not accounted for in this work) on the experimentally measured kink area [Fig. 3(c)] [9].

The experimental observation of a substantial doping dependence in the cuprate kink size had also been puzzling. From the underdoped towards overdoped regime, the measured apparent kink size monotonically decreases [3,21–23], which has commonly been interpreted as a weakening of the mode-coupling strength (for mechanisms with either phonons or spin fluctuations) [3,4,10,13,26,65,66]. Figure 4(a) plots the doping-dependent kink area data, where the *GW*PT trend also nicely agrees with experiments [3,23]. However, we discovered that this doping dependence in the kink size does not represent a doping dependence in the mode-coupling strength. As seen in Fig. 4(b), the magnitudes of the CuO-plane breathing-mode induced major peak in $\mathrm{Re}\Sigma_{n\mathbf{k}}^{e-\mathrm{ph}}$ (around 75 meV binding energy) at $x = 0.15$ and $x = 0.30$, are nearly identical from our *GW*PT calculations. This finding from our first-principles results is in stark contrast to the common interpretation, and invalidates the conclusion that the bosonic mode-coupling strength weakens with increased hole doping in the cuprates, at least for the phonons.

The observed significant doping dependence in the apparent kink size (as extracted with the commonly used procedure discussed above) is traced back to the tail height of the $\mathrm{Re}\Sigma_{n\mathbf{k}}^{e-\mathrm{ph}}$ at binding energies beyond $\omega_{\mathrm{ph}}^{\mathrm{max}}$. Specifically at $E_1$, $\mathrm{Re}\Sigma_{n\mathbf{k}_1}^{e-\mathrm{ph}}(E_1)$ differs dramatically at the two doping levels [5.3 meV for $x = 0.15$ and 12.8 meV for $x = 0.30$, see Fig. 4(b)], resulting in disparate slopes for the reference line drawn from $E_F$ to $E_1$. Note that the kink is defined as the deviation from the reference line; even if the peak values in $\mathrm{Re}\Sigma_{n\mathbf{k}}^{e-\mathrm{ph}}$ at two different doping levels are alike, the different slopes of the reference lines can give different apparent kink sizes. The difference in the tail height $\mathrm{Re}\Sigma_{n\mathbf{k}_1}^{e-\mathrm{ph}}(E_1)$ at different doping levels is largely caused by features in the electron DOS distribution, reflecting the multiband nature of the cuprates. We demonstrate this effect with rigid-band approximation calculations: using the as-calculated electron band structure, phonon dispersions, *e*-ph matrix elements at $x = 0.15$, we recompute $\Sigma_{n\mathbf{k}}^{e-\mathrm{ph}}$ after rigidly shifting $E_F$ to the corresponding value at $x = 0.30$ (to isolate the DOS effects on the kinks). As shown in Fig. 4(b) (and Fig. S2), the shift of $E_F$ to higher hole doping level (without changing the *e*-ph matrix elements) significantly elevates the tail height in $\mathrm{Re}\Sigma_{n\mathbf{k}}^{e-\mathrm{ph}}$. This arises because the non-resonant contributions to $\mathrm{Re}\Sigma_{n\mathbf{k}_1}^{e-\mathrm{ph}}(E_1)$ are inversely correlated to the energy difference between $E_1$ and the final states (denoted as $\Delta E$). The calculated electron band structure and DOS [Figs. 4(c)–(e)] show that at $x =$ 0.30, the quasi-hole at $E_1$ has a larger amplitude to be non-resonantly scattered into the deeper final electronic states (smaller $\Delta E$ from $E_1$ to the rise of part of the DOS below –0.2 eV and to further deeper states) than at $x = 0.15$, resulting in a higher value for $\mathrm{Re}\Sigma_{n\mathbf{k}_1}^{e-\mathrm{ph}}(E_1)$, and thus a decreased apparent kink size (see Eq. (S7) in Supplemental Material [28]). We further note that although DFTP calculation also includes this multi-band DOS effect, its unrealistically small *e*-ph coupling values fail to capture the correct doping trend in the kink size [Fig. 4(a)].

In summary, first-principles *GW*PT calculations of LSCO have identified that the correlation-enhanced *e*-ph interaction is the main origin of the ubiquitous 70–80 meV dispersion kink in the ARPES of cuprates, and revealed that the multi-band DOS effect plays a requisite role in understanding the doping dependence of the extracted apparent kink size. The quantitative agreement with experiments indicates that *GW*PT satisfactorily captures the correlation effects in *e*-ph coupling matrix elements in optimally doped and overdoped cuprates, where the quasiparticle picture is still appropriate [68–70].

**Acknowledgments**

This work was supported by the Theory of Materials Program at the Lawrence Berkeley National Laboratory (LBNL) through the Office of Basic Energy Sciences, U.S. Department of Energy under Contract No. DE-AC02-05CH11231, which provided for the DFPT and *GW*PT calculations, and by the National Science Foundation under Grant No. DMR-1926004, which provided for the Wannier interpolation calculations and relevant analysis. Advanced codes were provided by the Center for Computational Study of Excited-State Phenomena in Energy Materials (C2SEPEM) at LBNL, which is funded by the U.S. Department of Energy, Office of Science, Basic Energy Sciences, Materials Sciences and Engineering Division under Contract No. DE-AC02-05CH11231, as part of the Computational Materials Sciences Program. An award of computer time was provided by the Innovative and Novel Computational Impact on Theory and Experiment (INCITE) program. Significant amount of computational resources were needed in this work, and were provided by the following computing centers: Stampede2 at the Texas Advanced Computing Center (TACC) through Extreme Science and Engineering Discovery Environment (XSEDE), which is supported by National Science Foundation under Grant No. ACI-1053575; Summit at the Oak Ridge Leadership Computing Facility through the INCITE program, which is a DOE Office of Science User Facility supported under Contract No. DE-AC05-00OR22725; Cori at National Energy Research Scientific Computing Center (NERSC), which is supported by the Office of Science of the U.S. Department of Energy under Contract No. DE-AC02-05CH11231; Frontera at TACC, which is supported by National Science Foundation under Grant No. OAC-1818253. Z.L. acknowledges Y. He for many fruitful discussions and for sharing experimental perspectives. The authors thank A. Lanzara, G. Antonius, F. da Jornada, M. Del Ben, C. Yang, J. Deslippe, T. Cao, Z.-X. Li, M. Yi, C. S. Ong, H. J. Choi, and S. Ju for helpful discussions.

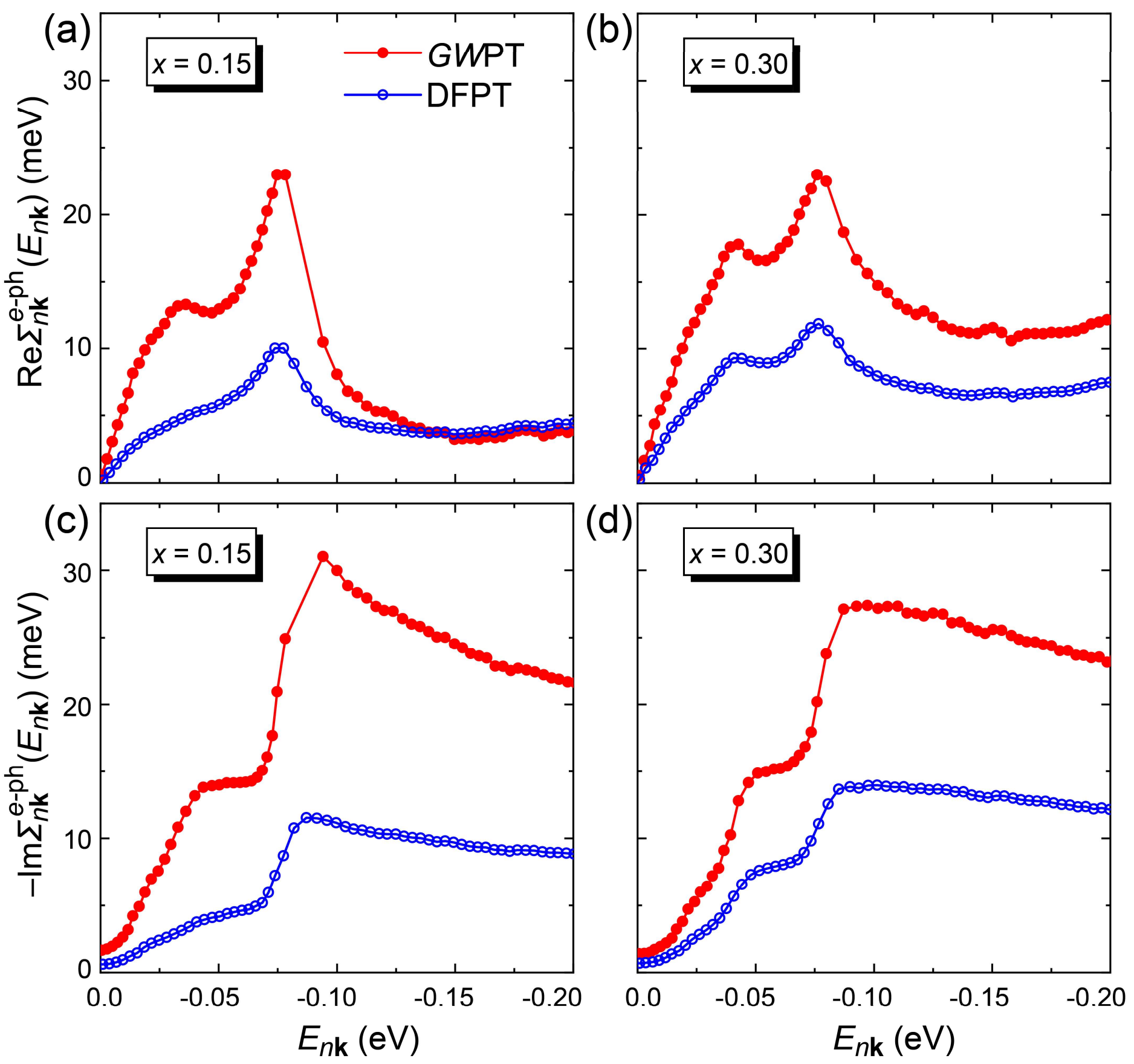

FIG. 1. (a),(b) Contribution of electron-phonon coupling to the real part of the nodal electron self-energy Re$\Sigma_{n\mathbf{k}}^{e-\mathrm{ph}}$ for (a) optimally doped ($x$ = 0.15) and (b) overdoped ($x$ = 0.30) LSCO at $T$ = 20 K, calculated using $e$-ph matrix elements from $GW$PT (red solid dots) and DFPT (blue empty dots). The Fermi energy $E_F$ is set to zero. (c),(d) Imaginary part of the phonon-induced contribution ($T$ = 20 K) to the electron self-energy Im$\Sigma_{n\mathbf{k}}^{e-\mathrm{ph}}$, the magnitude of which continuously rises until the maximum phonon frequency $\omega_{\mathrm{ph}}^{\max} \sim$ 87 meV.

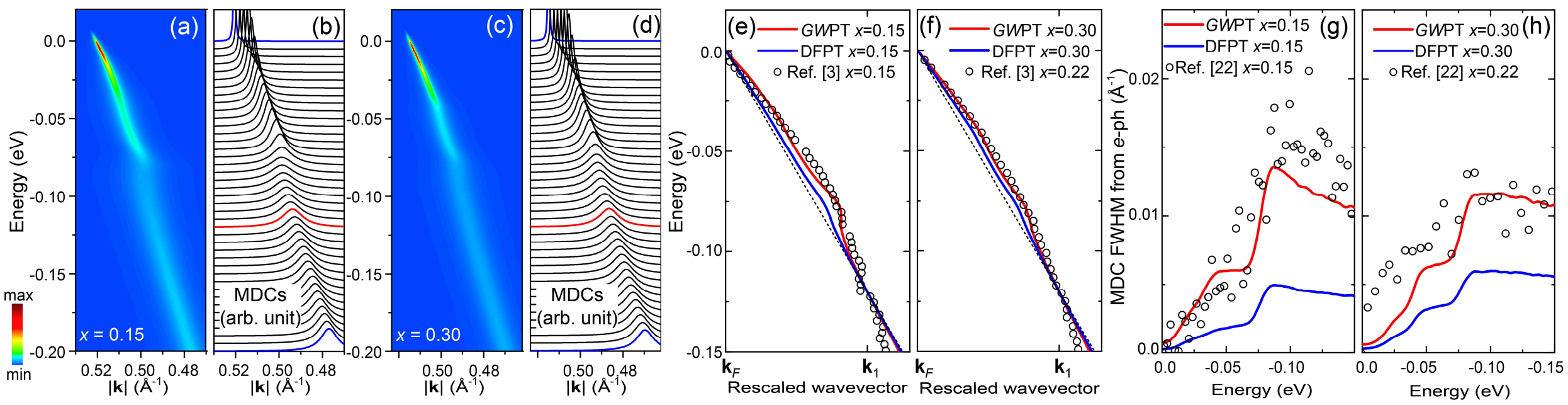

FIG. 2. (a) Spectral function $A_{n\mathbf{k}}(\omega)$ (plotted in color scale) calculated with $\Sigma_{n\mathbf{k}}^{e-\mathrm{ph}}(\omega)$ using *GW*PT for optimally doped ($x$ = 0.15) LSCO at $T$ = 20 K, along the nodal direction in the Brillouin zone. (b) Momentum-distribution curves (MDCs) at each energy slice between $E_F$ and –0.2 eV with 0.005 eV spacing. Each MDC is fitted to a Lorentzian function to get the peak position and linewidth. The MDCs at the extremal energies are indicated by the blue lines. The red line is the MDC at $E_1$ = –0.12 eV whose peak position is used to determine the bare-band straight reference line (connecting the peaks at $E_F$ and $E_1$ on a given dispersion relation). (e) Comparison of MDC-derived dispersion relation between experiment (open circles) [3] and theory from *GW*PT (red line) and DFPT (blue line), after aligning the reference line (black dashed line). $\mathbf{k}_F$ is the Fermi wavevector and $\mathbf{k}_1$ is the wavevector corresponding to $E_1$. (g) The $e$-ph contribution to the full-width-at-half-maximum (FWHM) of MDC peaks extracted from experiments (open circles) [22] and from *GW*PT (red line) and DFPT (blue line). (c),(d),(f),(h) Similar to (a),(b),(e),(g) but for overdoped LSCO.

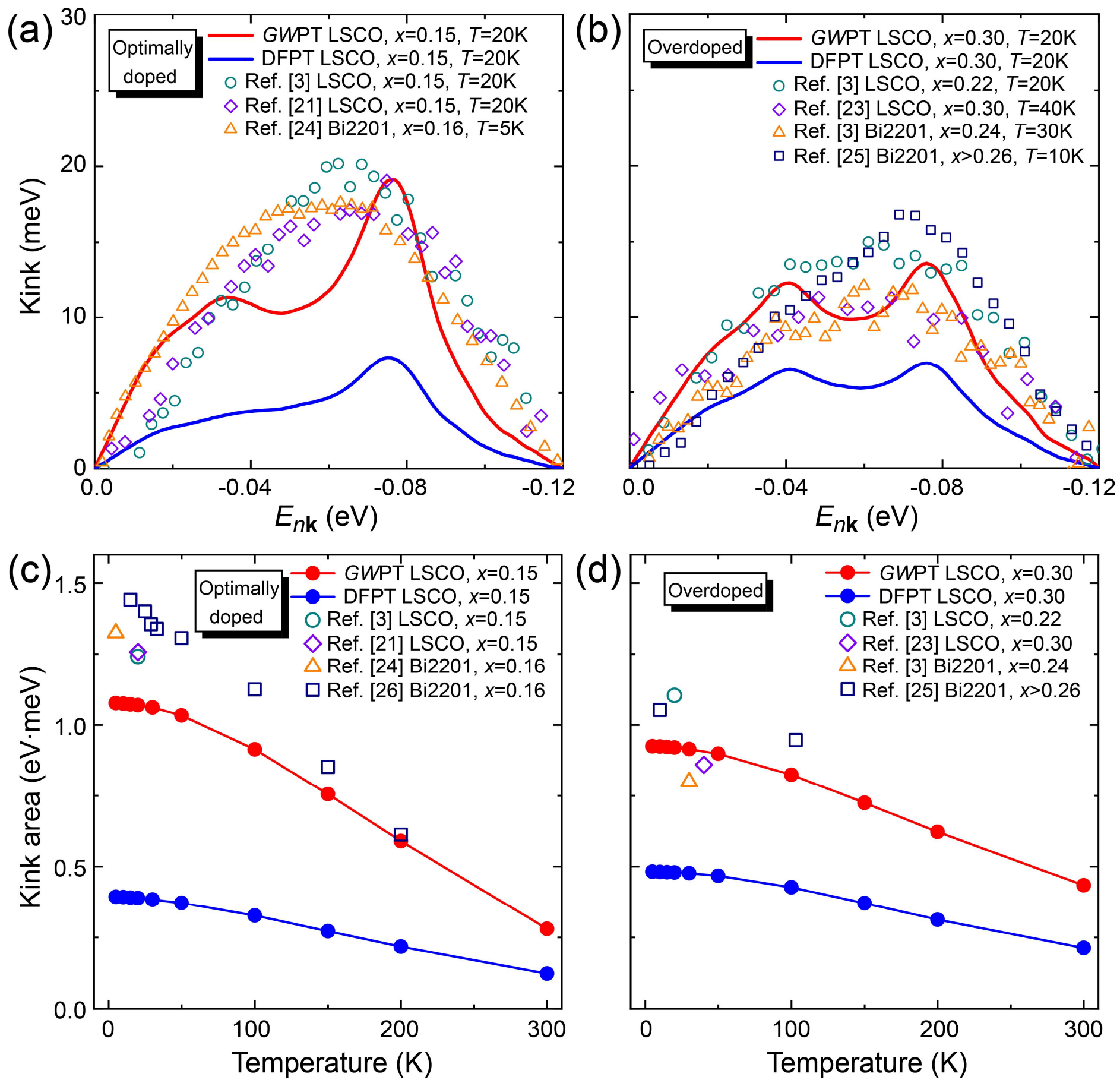

FIG. 3. (a),(b) Kink size as a function of quasiparticle-state energy, extracted as the deviation of the MDC-derived dispersion relation from the straight reference line (see Fig. 2). Data shown are from various experiments [3,21,23–26] on LSCO and Bi2201 (open symbols) and from the calculations of LSCO using *GW*PT (red dots) and DFPT (blue dots), at the optimally doped and overdoped regimes. (c),(d) Total kink area, defined as the area under the curve in (a) and (b), as a function of temperature.

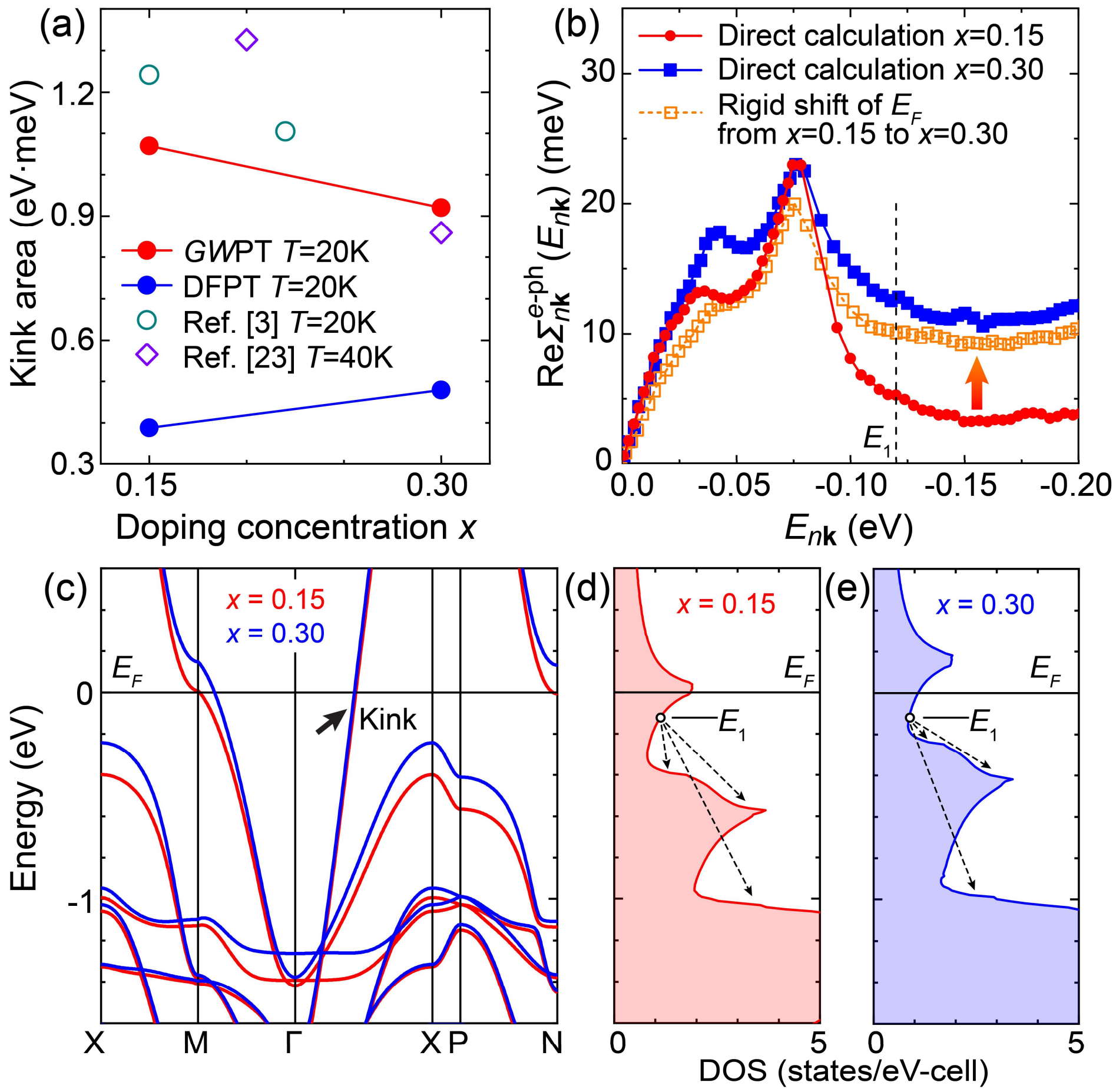

FIG. 4. (a) Doping dependence in the total apparent kink area of LSCO from experiments [3,23], *GW*PT, and DFPT calculations. (b) Direct calculations of $\mathrm{Re}\Sigma_{n\mathbf{k}}^{e-\mathrm{ph}}$ at $x$ = 0.15 (red solid dot) and $x$ = 0.30 (blue solid squares) at $T$ = 20 K. The orange empty squares represent a rigid-band approximation calculation with a rigid shift of $E_F$ from that of $x$ = 0.15 to that of $x$ = 0.30 in constructing $\Sigma_{n\mathbf{k}}^{e-\mathrm{ph}}$ (i.e. the band structure, phonon dispersions, and $e$-ph matrix elements remain unchanged at the $x$ = 0.15 level), resulting in the elevation of the self-energy tail, as indicated by the arrow. (c) Band structure of LSCO at $x$ = 0.15 and $x$ = 0.30. The black arrow points to the position of the kink along the Γ−X nodal direction. (d),(e) DOS of LSCO at the indicated two doping levels. The scatterings (indicated by arrow-headed dashed lines) of quasi-hole from $E_1$ to higher-binding-energy hole final states with significant DOS have larger energy separations (thus contributing less to $\mathrm{Re}\Sigma_{n\mathbf{k}_1}^{e-\mathrm{ph}}(E_1)$) at $x$ = 0.15 than at $x$ = 0.30.

Supplemental Material for "Unmasking the Origin of Kinks in the Photoemission Spectra of Cuprate Superconductors"

Zhenglu Li, Meng Wu, Yang-Hao Chan, and Steven G. Louie*

*Department of Physics, University of California at Berkeley, Berkeley, CA 94720, USA and Materials Sciences Division, Lawrence Berkeley National Laboratory, Berkeley, CA 94720, USA.*

*Email: sglouie@berkeley.edu

## I. Computational details

This work involves three major computation steps: 1) DFPT calculation using the ABINIT code [29]; 2) *GW*PT calculation using the BerkeleyGW code [19,20,30,31]; and 3) Wannier interpolation using the EPW code [32,59]. We have developed wrappers to link the three codes.

We perform DFT and DFPT calculations of $La_{2-x}Sr_xCuO_4$ (LSCO) using the ABINIT code [29]. The Sr doping effect of $x$ = 0.15 and $x$ = 0.30 is simulated by removing electrons from the crystal and adding a compensating negative charge background [5]. We use norm-conserving pseudopotentials [33,34] with generalized gradient approximation to the exchange-correlation functional [35], and a plane-wave basis with a cutoff set to 100 Ry. At each doping, the structures are fully optimized by minimizing the forces on each of the 7 atoms in the primitive cell of the tetragonal crystal structure (the primitive cell lattice vectors are $\mathbf{a}_1 = (0, \frac{a}{\sqrt{2}}, -\frac{c}{2})$, $\mathbf{a}_2 = (0, -\frac{a}{\sqrt{2}}, -\frac{c}{2})$, $\mathbf{a}_3 = (-\frac{a}{\sqrt{2}}, 0, \frac{c}{2})$; at $x$ = 0.15, the optimized $a$ = 3.777 Å and $c$ = 13.106 Å; at $x$ = 0.30, the optimized $a$ = 3.750 Å and $c$ = 12.981 Å), followed by DFT (**k** grid set to 32×32×32) and DFPT (**k** grid set to 8×8×8 and **q** grid set to 4×4×4) calculations for the ground-state properties and the linear-response changes due to atom displacements (21 patterns) at each **q**, respectively.

The *GW*PT calculation is performed with a screened Coulomb interaction cutoff of 25 Ry and a bare Coulomb interaction cutoff of 100 Ry using the BerkeleyGW code [20,30,31]. The inverse dielectric matrix $\epsilon^{-1}_{\mathbf{GG'}}(\mathbf{p})$ (where **G**'s are the reciprocal lattice vectors) and the linear-response changes in the self-energy $\partial_{\mathbf{q}\nu}\Sigma^{e-e}(\mathbf{r}, \mathbf{r}'; \varepsilon)$ are constructed with 200 bands and 8×8×8 internal **p** grid [20].

The $e$-ph matrix elements $g^{\mathrm{DFT}}_{mn\nu}(\mathbf{k}, \mathbf{q})$ and $g^{GW}_{mn\nu}(\mathbf{k}, \mathbf{q})$ from DFPT and *GW*PT, respectively, are directly computed on coarse 4×4×4 **k** and **q** grids, and then interpolated with a Wannier function approach [36,37] using the EPW code [32]. The gauge of the wavefunctions and matrix elements must be handled consistently through the whole procedure (i.e., in DFPT, *GW*PT, and Wannier interpolation). The Wannier subspace includes 17 Wannier orbitals corresponding to the five $d$ orbitals of the Cu atom and three $p$ orbitals of each of the four O atoms. To compute the phonon-induced electron self-energy $\Sigma^{e-\mathrm{ph}}_{n\mathbf{k}}$, 1,000,000 random phonon **q** points are sampled along with a 2 meV smearing factor. For the calculation of the $e$-ph coupling constant $\lambda$, interpolated 32×32×32 **k** and 16×16×16 **q** grids are used. Note that the computation of $\Sigma^{e-\mathrm{ph}}_{n\mathbf{k}}$ requires a much denser sampling to resolve the fine energy-dependent features within the phonon frequency range, whereas $\lambda$ is an averaged quantity on the Fermi surface and therefore converges relatively faster. Other convergence parameters are consistent with the previous DFPT-level study [5] and are tested by ourselves to produce converged results.

The DFPT calculations directly compute all the $e$-ph matrix elements (full $m$, $n$, $\nu$, **k**, and **q** dependence) within the Cu-O manifolds. However, because of the extremely expensive computational resource required by *GW*PT calculations, we compute a subset of *GW*PT matrix elements that gives the most dominant contributions. The *GW*PT calculations directly compute (for all phonon $\mathbf{q}\nu$ modes): 1) the matrix elements within the single band crossing $E_F$ among all **k** and **q** points, and 2) the matrix elements from the single coarse-grid $\mathbf{k}$ point on the band crossing $E_F$ (the electron state closest to the nodal kink on the coarse grid) to all the other 16 Cu-O bands via all $\mathbf{q}$ points. The less important matrix elements are replaced by their DFPT values in the following Wannier interpolation calculations. We have checked that this subset of explicitly computed *GW*PT *e*-ph matrix elements accounts for the overwhelming majority of the correlation enhancement effect and produces the main physics of the photoemission kink. Our procedure in fact places a lower bound on the correlation enhancement of the *e*-ph coupling.

Note that soft phonon modes may arise from direct $T = 0$ K DFPT calculations because we are forcing a tetragonal primitive unit cell whereas in reality structural distortions set in at low temperature [5]. We have thus adopted a common phonon stabilization procedure using high-temperature sampled structures to proceed with the *e*-ph coupling calculations [20,38,39]. We perform *ab initio* molecular dynamics calculations with 2×2×2 supercells at 300 K using the VASP code [40] with projected augmented wave pseudopotentials [41]. After thermal equilibrium is reached, 50 randomly sampled structures are picked with their forces being fitted to harmonic potentials using the ALAMODE code [42]. The dynamical matrices are constructed using the fitted harmonic potentials and then later read by EPW for interpolations.

## II. Theoretical formalism

The core quantity for building microscopic *e*-ph theories is the *e*-ph matrix element $g_{mn\nu}(\mathbf{k},\mathbf{q})$ that measures the coupling amplitude of a quasiparticle from the initial quasiparticle state $|\psi_{n\mathbf{k}}\rangle$ to a final state $|\psi_{m\mathbf{k}+\mathbf{q}}\rangle$ via a phonon mode labeled by $\mathbf{q}\nu$. Such a process can be directly and efficiently calculated using first principles in the primitive unit cell by taking advantage of linear-response theory. Within *GW*PT, the *e*-ph matrix elements at the *GW* level are constructed as [20],

$$g_{mn\nu}^{GW}(\mathbf{k},\mathbf{q}) = \langle\psi_{n\mathbf{k}}|\partial_{\mathbf{q}\nu}V_{\text{ion}} + \partial_{\mathbf{q}\nu}V_{\text{H}}|\psi_{m\mathbf{k}+\mathbf{q}}\rangle + \langle\psi_{n\mathbf{k}}|\partial_{\mathbf{q}\nu}\Sigma^{e-e}(\mathbf{r},\mathbf{r}';\varepsilon)|\psi_{m\mathbf{k}+\mathbf{q}}\rangle, \tag{S1}$$

where $V_{\text{ion}}$ is the ionic potential seen by the electrons, $V_{\text{H}}$ is the Hartree potential, and $\Sigma^{e-e}(\mathbf{r},\mathbf{r}';\varepsilon)$ is the electron self-energy operator that contains the electron-electron interaction effects. The *e*-ph matrix element at the DFT level $g_{mn\nu}^{\text{DFT}}(\mathbf{k},\mathbf{q})$ is calculated using DFPT in which $\partial_{\mathbf{q}\nu}\Sigma^{e-e}(\mathbf{r},\mathbf{r}';\varepsilon)$ is simply approximated as the change in the static exchange-correlation potential $\partial_{\mathbf{q}\nu}V_{\text{xc}}(\mathbf{r})$. The resulting matrix elements in fact only describe the interaction of the fictitious independent Kohn-Sham particles with the phonons. A calculation of the linear-response change in $\Sigma^{e-e}$ due to phonons using *GW*PT captures physically the many-electron self-energy effects in the interaction of a true quasiparticle with phonons (i.e., the true *e*-ph matrix element $g_{mn\nu}^{GW}(\mathbf{k},\mathbf{q})$), hence providing more accurate descriptions of the *e*-ph interactions. The change in the *GW* self-energy is [20],

$$\partial_{\mathbf{q}\nu}\Sigma^{e-e}(\mathbf{r},\mathbf{r}';\varepsilon) = i\int\frac{d\varepsilon'}{2\pi}e^{-i\delta\varepsilon'}\partial_{\mathbf{q}\nu}G(\mathbf{r},\mathbf{r}';\varepsilon-\varepsilon')W(\mathbf{r},\mathbf{r}';\varepsilon'), \tag{S2}$$

where $\delta = 0^+$, $G(\mathbf{r},\mathbf{r}';\varepsilon-\varepsilon')$ is the single-particle Green's function, and $W(\mathbf{r},\mathbf{r}';\varepsilon')$ is the screened Coulomb interaction. In Eq. (S2), we neglected the change in $W$ due to phonons by adopting the constant-screening approximation [20,43], which is equivalent to the well-justified approximation $\frac{\delta W}{\delta G} \approx 0$ used in the *ab initio GW* plus Bethe-Salpeter equation (*GW*-BSE) formalism for exciton physics [44–46].

The *e*-ph interaction contribution to the electron self-energy $\Sigma^{e-\text{ph}}$ can then be constructed at either the *GW*PT level (using $g_{mn\nu}^{GW}(\mathbf{k},\mathbf{q})$) or at the DFPT level (using $g_{mn\nu}^{\text{DFT}}(\mathbf{k},\mathbf{q})$) within the Fan-Migdal approximation as [17,60–62,32],

$$\Sigma_{n\mathbf{k}}^{e-\text{ph}}(\omega) = \sum_{\mathbf{q}\nu}\sum_{m}|g_{mn\nu}(\mathbf{k},\mathbf{q})|^2 \times \left[\frac{f_{m\mathbf{k}+\mathbf{q}} + n_{\mathbf{q}\nu}}{\omega - \varepsilon_{m\mathbf{k}+\mathbf{q}} + \omega_{\mathbf{q}\nu} + i\delta} + \frac{1 - f_{m\mathbf{k}+\mathbf{q}} + n_{\mathbf{q}\nu}}{\omega - \varepsilon_{m\mathbf{k}+\mathbf{q}} - \omega_{\mathbf{q}\nu} + i\delta}\right], \tag{S3}$$

where $f_{n\mathbf{k}}(T)$ is the Fermion occupation number and $n_{\mathbf{q}\nu}(T)$ is the Boson occupation number, $\varepsilon_{n\mathbf{k}}$ is the DFT eigenvalue, and $\omega_{\mathbf{q}\nu}$ is the phonon frequency. Self-consistently using $GW$ eigenvalues, which undergo band reordering from the DFT results within the Cu-O bands, for Eq. (S3) is technically challenging when combined with Wannier interpolation. However, this may be avoided since the effect from $GW$ correction in the band structure (via Fermi velocity and band width) to Eq. (S3) is estimated to be one order of magnitude smaller than the effect from the $GW$ renormalization in the $e$-ph coupling strength. The first (second) term in the bracket corresponds to phonon emission (absorption) processes. At low temperature, the phonon emission process dominates $\Sigma^{e-\mathrm{ph}}$ and hence the phonon-induced kink in the photoemission spectrum. $\mathrm{Re}\Sigma_{n\mathbf{k}}^{e-\mathrm{ph}}(E_F)$ is subtracted off from the computed $\Sigma_{n\mathbf{k}}^{e-\mathrm{ph}}(\omega)$ to impose the sum rule of electron number conservation [17,32]. The electron spectral function including interaction with phonons is constructed from the calculated phonon-induced part of the self-energy as [17,32],

$$A_{n\mathbf{k}}(\omega) = \frac{1}{\pi}\frac{\left|\mathrm{Im}\Sigma_{n\mathbf{k}}^{e-\mathrm{ph}}(\omega)\right|}{\left|\omega - \varepsilon_{n\mathbf{k}} - \mathrm{Re}\Sigma_{n\mathbf{k}}^{e-\mathrm{ph}}(\omega)\right|^2 + \left|\mathrm{Im}\Sigma_{n\mathbf{k}}^{e-\mathrm{ph}}(\omega)\right|^2}, \quad \text{(S4)}$$

and the Fermi factor $f_{n\mathbf{k}}(T)$ is imposed in plotting the intensity map [67] of Figs. 2(a) and 2(c) in the main text.

## III. Low-energy behavior of phonon-induced part of self-energy

Figure 1 in the main text shows a rapid variation in $\mathrm{Re}\Sigma_{n\mathbf{k}}^{e-\mathrm{p}}$ and $\mathrm{Im}\Sigma_{n\mathbf{k}}^{e-\mathrm{ph}}$ mainly occurs for electronic states between $E_F$ and a binding energy corresponding to the maximum phonon frequency $\omega_{\mathrm{ph}}^{\max}$ ~ 87 meV in LSCO. It is well-understood within the $e$-ph mechanism. At low temperature, a quasi-hole state $n\mathbf{k}$ excited below $E_F$ can scatter to other hole states via emitting a phonon, and two resonant scattering regimes (along with non-resonant scatterings) contribute to $\Sigma_{n\mathbf{k}}^{e-\mathrm{ph}}(E_{n\mathbf{k}})$: 1) If the hole excitation energy (given by $E_F - E_{n\mathbf{k}}$) is within the range of all possible phonon frequencies, i.e. $E_F - E_{n\mathbf{k}} < \omega_{\mathrm{ph}}^{\max}$, the final hole states must have a yet still positive but lower excitation energy (i.e. smaller binding energy in ARPES) than $E_F - E_{n\mathbf{k}}$ (energy conservation via emitting a phonon), restricting some phonon modes from participating. Therefore, $\Sigma_{n\mathbf{k}}^{e-\mathrm{ph}}(E_{n\mathbf{k}})$ is expected to be finely structured through phonon-mode dependent coupling strength as well as the phonon DOS. 2) In contrast, if $E_F - E_{n\mathbf{k}} > \omega_{\mathrm{ph}}^{\max}$, all phonon channels are available for emission, and thus the scattering process is dominated by the final hole DOS, making $\Sigma_{n\mathbf{k}}^{e-\mathrm{ph}}(E_{n\mathbf{k}})$ a smoother function of energy [5]. The resonant channels uniquely determine the scattering rates reflected in $\mathrm{Im}\Sigma_{n\mathbf{k}}^{e-\mathrm{p}}$, whereas $\mathrm{Re}\Sigma_{n\mathbf{k}}^{e-\mathrm{ph}}$ has contributions from both resonant and non-resonant processes.

In a simple Holstein model where a single electron band couples to an Einstein phonon mode, the peak of the $\mathrm{Re}\Sigma_{n\mathbf{k}}^{e-\mathrm{ph}}$ is determined by a single uniform $e$-ph coupling parameter $|g|$. However, in complex materials (such as LSCO in this work), the $e$-ph mode-coupling strength reflected in $\mathrm{Re}\Sigma_{n\mathbf{k}}^{e-\mathrm{ph}}$ at a particular phonon mode frequency involves contributions from all other electronic states [see Eq. (S3)].

## IV. Extraction of kinks from spectral functions

The procedure for the extraction of the size and position of the photoemission kinks is described in the main text, following the common experimental practice. We note that the DFT band structure [Fig. 4(c) in the main text] within the energy range of interest (along the nodal direction) is well represented as a linear dispersion relation $\varepsilon_{n\mathbf{k}} = \hbar\mathbf{v}_0 \cdot (\mathbf{k} - \mathbf{k}_F)$ where $\mathbf{v}_0$ is the DFT Fermi velocity (note that $E_F$ is set to zero here). The quasiparticle energy due to *e*-ph interaction at a given temperature is,

$$E_{n\mathbf{k}} = \varepsilon_{n\mathbf{k}} + \mathrm{Re}\Sigma_{n\mathbf{k}}^{e-\mathrm{ph}}(E_{n\mathbf{k}}). \tag{S5}$$

Since the reference line is determined by the two points at $\mathbf{k}_F$ and $\mathbf{k}_1$ on the dispersion relation along the nodal direction, we can write the expression of this reference line [black dashed lines in Figs. 2(e) and 2(f) in the main text] as,

$$e_{n\mathbf{k}} = E_1 \times \frac{|\mathbf{k} - \mathbf{k}_F|}{|\mathbf{k}_1 - \mathbf{k}_F|}. \tag{S6}$$

The kink is extracted as the deviation of $E_{n\mathbf{k}}$ from $e_{n\mathbf{k}}$ at each $\mathbf{k}$, i.e., $\mathrm{kink}(E_{n\mathbf{k}}) = E_{n\mathbf{k}} - e_{n\mathbf{k}}$, and by using the linearity of $\varepsilon_{n\mathbf{k}}$, it becomes,

$$\mathrm{kink}(E_{n\mathbf{k}}) = \mathrm{Re}\Sigma_{n\mathbf{k}}^{e-\mathrm{ph}}(E_{n\mathbf{k}}) - \mathrm{Re}\Sigma_{n\mathbf{k}_1}^{e-\mathrm{ph}}(E_1) \times \frac{|\mathbf{k} - \mathbf{k}_F|}{|\mathbf{k}_1 - \mathbf{k}_F|}, \quad \text{with } E_{n\mathbf{k}} \in [E_F, E_1]. \tag{S7}$$

Eq. (S7) is true in the quasiparticle limit ($\mathrm{Im}\Sigma \ll$ band width). Moreover, when $\mathrm{Re}\Sigma_{n\mathbf{k}}^{e-\mathrm{ph}}$ (which is 20 – 30 meV in peak value) is comparatively small to the electron band energy range of interest (100 – 200 meV in this work), the kink strength can be *approximately* expressed as,

$$\mathrm{kink}(E_{n\mathbf{k}}) \approx \mathrm{Re}\Sigma_{n\mathbf{k}}^{e-\mathrm{ph}}(E_{n\mathbf{k}}) - \mathrm{Re}\Sigma_{n\mathbf{k}_1}^{e-\mathrm{ph}}(E_1) \times \frac{E_{n\mathbf{k}}}{E_1}. \tag{S8}$$

Eq. (S8) means that one can draw a straight reference line connecting the two points, $(E_F, \mathrm{Re}\Sigma_{n\mathbf{k}_F}^{e-\mathrm{ph}}(E_F))$ (i.e. the origin) and $(E_1, \mathrm{Re}\Sigma_{n\mathbf{k}_1}^{e-\mathrm{ph}}(E_1))$ on the $\mathrm{Re}\Sigma_{n\mathbf{k}}^{e-\mathrm{ph}}$-*vs.*-$E_{n\mathbf{k}}$ curve, and the kink size can be *approximated* as the difference between $\mathrm{Re}\Sigma_{n\mathbf{k}}^{e-\mathrm{ph}}(E_{n\mathbf{k}})$ and this reference line at each $E_{n\mathbf{k}}$.

To extract the experimental kink data, we fit linearly data points with binding energy between 0 and 30 meV to determine $\mathbf{k}_F$. We fit linearly data points *available* (not all experiments considered span this full range) between 100 and 200 meV binding energies and use $E_1 = -0.12$ eV in the fitted dispersion relation to determine $\mathbf{k}_1$. Then we use the two points ($\mathbf{k}_F$, $E_F$) and ($\mathbf{k}_1$, $E_1$) to determine the reference bare-band line for the extraction of the kink deviation. The extracted kink data are consistent among different experiments [3,21,23–26], thus providing confident quantitative comparisons with theory.

The experimental MDC width is in general composed of several contributions, including impurity scattering (presumably relatively constant with respect to energy), electron-electron interaction (power law behavior as a function of binding energy), and *e*-ph interaction (saturates at energies beyond $\omega_{\mathrm{ph}}^{\max}$) [68]. Following standard practice by experimental groups, we extract the experimental *e*-ph MDC width plotted in Figs. 2(g) and 2(h), by first fitting a power law function to the original measured MDC FWHM data [22] between the binding energy of 100 and 200 meV for electron-electron contribution, and then subtracting off this fitted function and the impurity contribution (taken to be the width at $E_F$) from the original data in the whole range. The data points after subtraction are denoted as experimental MDC FWHM width from *e*-ph coupling and are plotted in Figs. 2(g) and 2(h) in the main text. Note that large uncertainty exists in the *e*-ph MDC width extraction, depending on the choice of fitting range, and therefore the comparisons in Figs. 2(g) and 2(h) are qualitative.

In some early experiments [21,22], the heavily overdoped ($x$ = 0.30) data show unsatisfactory statistics and the extracted quantity does not qualitatively agree with more recent data [23]. Therefore, we have not used those $x$ = 0.30 experimental data in comparisons with theory in Figs. 2–4 in the main text.

## V. Electron-phonon coupling constant $\lambda$

We calculate the *e*-ph coupling constant $\lambda$ on the Fermi surface [17,60–62], although phonons alone are not expected to be responsible for the *d*-wave superconductivity in single-layer cuprates [68]. Our *GW*PT calculation gives $\lambda = 0.55$ for $x = 0.15$, and $\lambda = 0.54$ for $x = 0.30$ (whereas DFPT gives $\lambda = 0.26$ for $x = 0.15$ and $\lambda = 0.27$ for $x = 0.30$, consistent with previous DFPT studies [5,6]), evidencing a moderately strong *e*-ph coupling constant in LSCO.

## VI. *GW*PT validation set of metals

To demonstrate the validity of the *GW*PT method and to highlight correlation-enhancement in the *e*-ph coupling in LSCO (and also in $Ba_{1-x}K_xBiO_3$ (BKBO) [20]), we performed a series of *GW*PT calculations of the *e*-ph coupling constant $\lambda$ on five conventional metals: 1) Cu, 2) Al, 3) In, 4) ZrN, and 5) Nb. This set of materials covers a non-superconductor, as well as weakly-, moderately- and strongly-coupled phonon-driven superconductors. For this validation set, the *e*-ph matrix elements at both DFPT and *GW*PT levels are computed on 8×8×8 **k** and 4×4×4 **q** grids which are then interpolated to finer grids by Wannier functions [32,36,37,59]. The DFPT calculations use a 60 Ry plane-wave cutoff. The *GW*PT calculations use a 15 Ry plane-wave cutoff and an 8×8×8 **p** grid for the screened Coulomb interaction, and a total of 200 bands for band summations. A dense Brillouin zone wavevector sampling may be needed for complicated Fermi surfaces composed of multiple bands with disconnected small pockets, such as in Al. The Wannierization processes [32,59] interpolate the *e*-ph matrix elements to finer 40×40×40 **k** and 20×20×20 **q** grids for the *e*-ph coupling calculations. Here we list the *initial* Wannier projections used for each material: 1) Cu: 3d, 4s, 4p; 2) Al: 3s, 3p, 3d; 3) In: 5s, 5p; 4) ZrN: 4s, 4p, 4d, 5s, 5p for Zr, and 2s, 2p for N; and 5) Nb: 4d, 5s, 5p. The *e*-ph coupling constant $\lambda$ is computed at both DFPT and *GW*PT levels, and are plotted in Fig. S3 together with results of BKBO from an earlier work [20]. The corresponding measured $\lambda$ from tunneling experiments [47–53] are included for comparison.

In this validation set, our DFPT results are consistent with other earlier DFPT calculations [54–57]. Given the experimental uncertainties (which may be up to few tens of percent, see discussions in, e.g. Ref. [58]) and the numerical uncertainties, we consider the overall agreement between the values for $\lambda$ from *GW*PT and experiments is good (maximum difference in $\lambda$ is ~ 0.1 for the five conventional metals). In the oxide BKBO, $\lambda$ is significantly enhanced by over a factor of 2 in *GW*PT (due to many-electron correlations) compared with DFPT (see Fig. S3 and Ref. [20]). Around the optimal doping of BKBO, the tunneling experiment extracted $\lambda^{\text{expt.}} = 1.2$ [53] and our previous *GW*PT results give $\lambda^{GW\text{PT}} = 1.14$ agreeing with the experiment, whereas DFPT fails significantly giving $\lambda^{\text{DFPT}} = 0.47$ [20]. This *GW*PT validation set on five conventional metals as well as an oxide metal establishes *GW*PT as a reliable method to study *e*-ph interaction and its use in exploring the intriguing interplay between many-electron correlations and *e*-ph interactions in more “unconventional” materials.

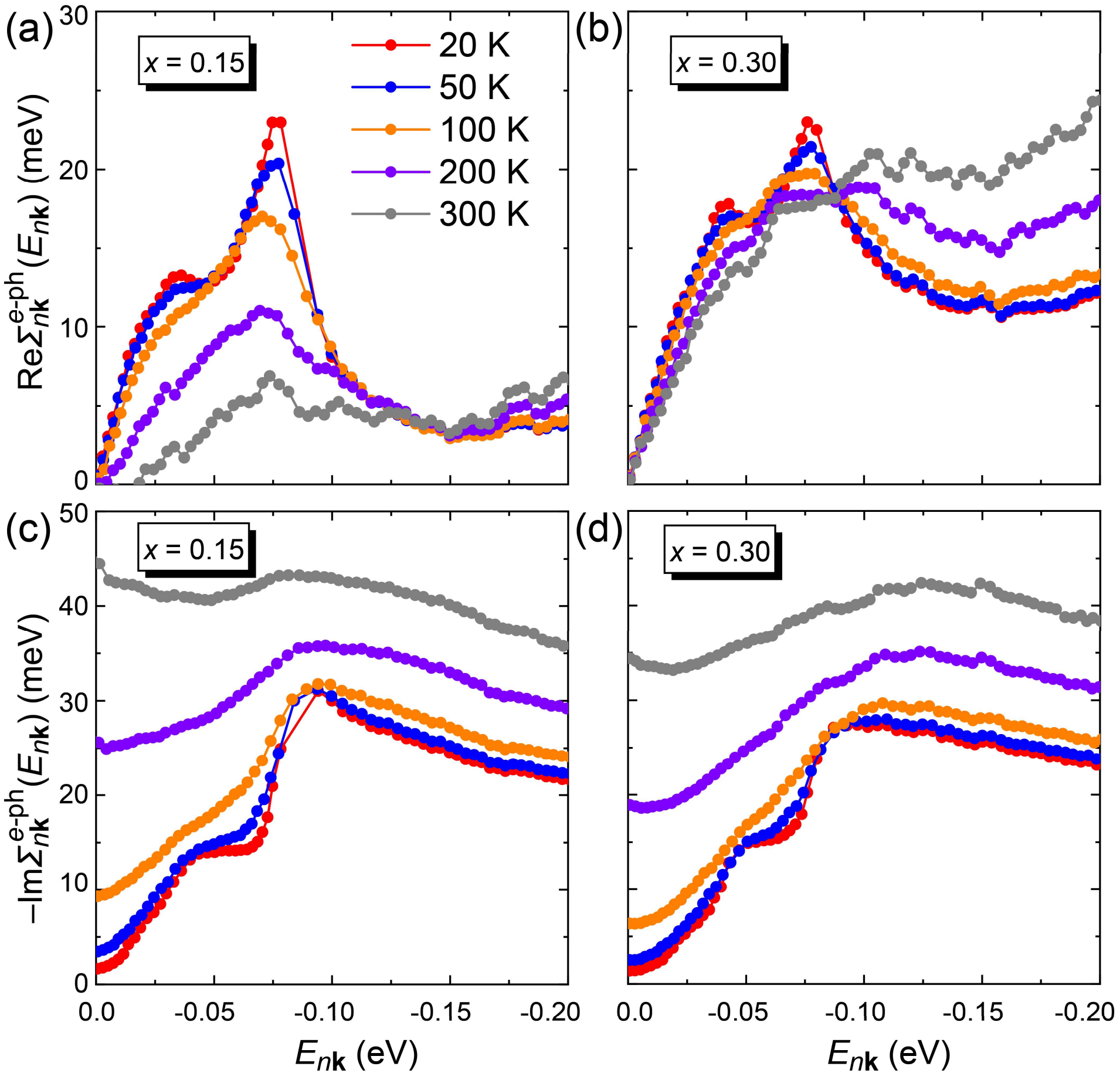

FIG. S1. (a),(b) Temperature dependence of $\mathrm{Re}\Sigma_{n\mathbf{k}}^{e-\mathrm{ph}}(E_{n\mathbf{k}})$ of LSCO at (a) $x = 0.15$ and (b) $x = 0.30$ from *GW*PT calculations. In the optimally doped case ($x$ = 0.15), increasing temperature reduces the $\mathrm{Re}\Sigma_{n\mathbf{k}}^{e-\mathrm{ph}}(E_{n\mathbf{k}})$ peak height, whereas in the overdoped case ($x = 0.30$), the high binding energy (> 0.1 eV) tail rises with increasing temperature. This is due to a closer proximity of the states to the higher density features of the DOS distribution in the overdoped case. (c),(d) Temperature dependence of $\mathrm{Im}\Sigma_{n\mathbf{k}}^{e-\mathrm{ph}}(E_{n\mathbf{k}})$ of LSCO at (c) $x = 0.15$ and (d) $x = 0.30$. The effects of high temperatures in both $\mathrm{Re}\Sigma_{n\mathbf{k}}^{e-\mathrm{ph}}$ and $\mathrm{Im}\Sigma_{n\mathbf{k}}^{e-\mathrm{ph}}$ blur the kink feature and broaden the spectral functions.

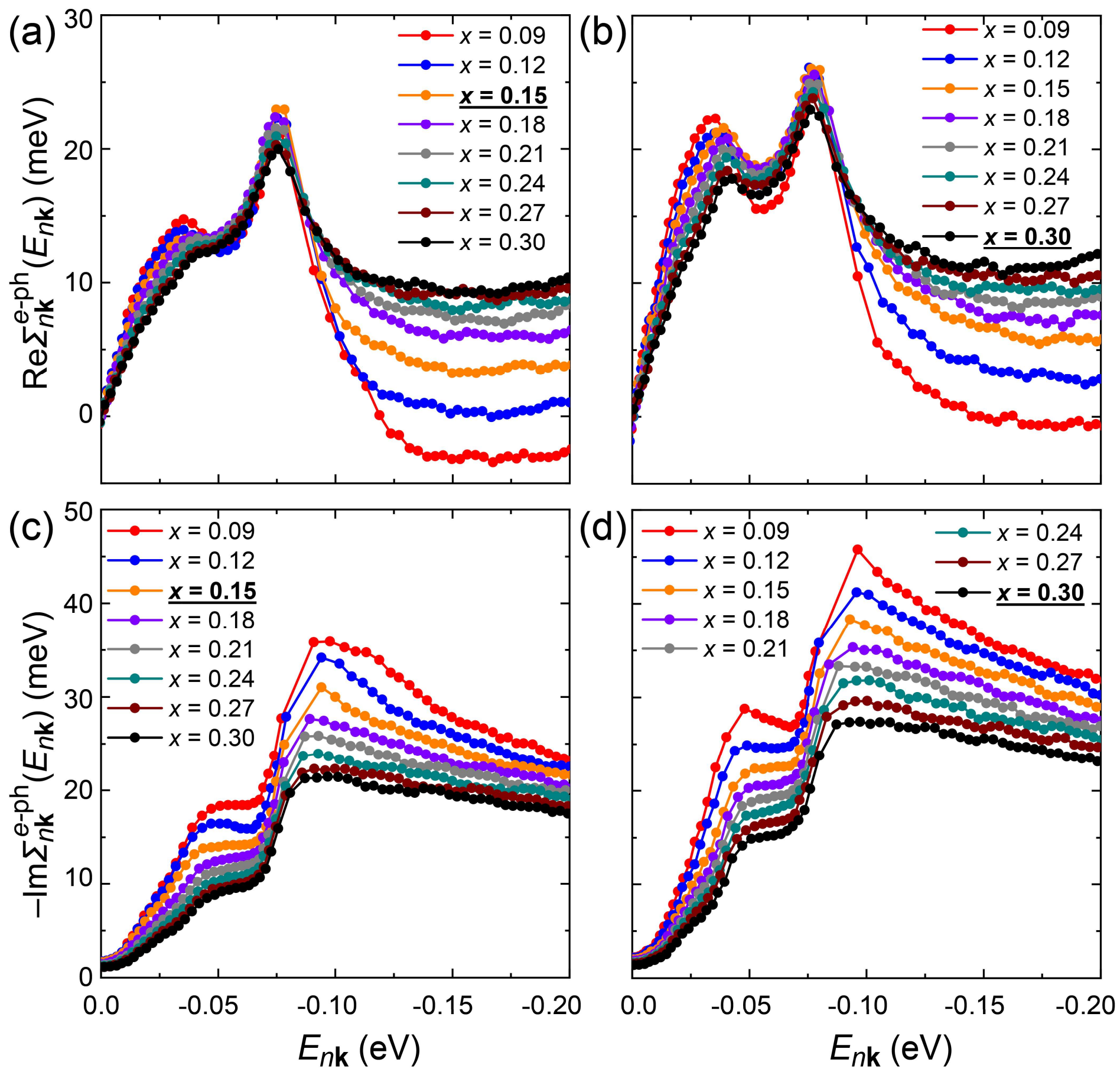

FIG. S2. (a),(b) Doping dependence of $\mathrm{Re}\Sigma_{n\mathbf{k}}^{e-\mathrm{ph}}(E_{n\mathbf{k}})$ (at *GW*PT level) constructed within rigid-band approximation using the *e*-ph matrix elements directly calculated at $x$ = 0.15 and $x$ = 0.30 (indicated by underline), given respectively by panels (a) and (b). A significant elevation of the high binding energy tail (> 0.1 eV) in the $\mathrm{Re}\Sigma_{n\mathbf{k}}^{e-\mathrm{ph}}(E_{n\mathbf{k}})$ with increasing hole doping is observed, due to the sharp multi-band features in the electron DOS. The main peak height of $\mathrm{Re}\Sigma_{n\mathbf{k}}^{e-\mathrm{ph}}(E_{n\mathbf{k}})$ shows only slight changes. (c),(d) Doping dependence of $\mathrm{Im}\Sigma_{n\mathbf{k}}^{e-\mathrm{ph}}(E_{n\mathbf{k}})$ (at *GW*PT level) constructed within rigid-band approximation using the *e*-ph matrix elements directly calculated at $x$ = 0.15 and $x$ = 0.30 (indicated by underline), given respectively in panels (c) and (d).

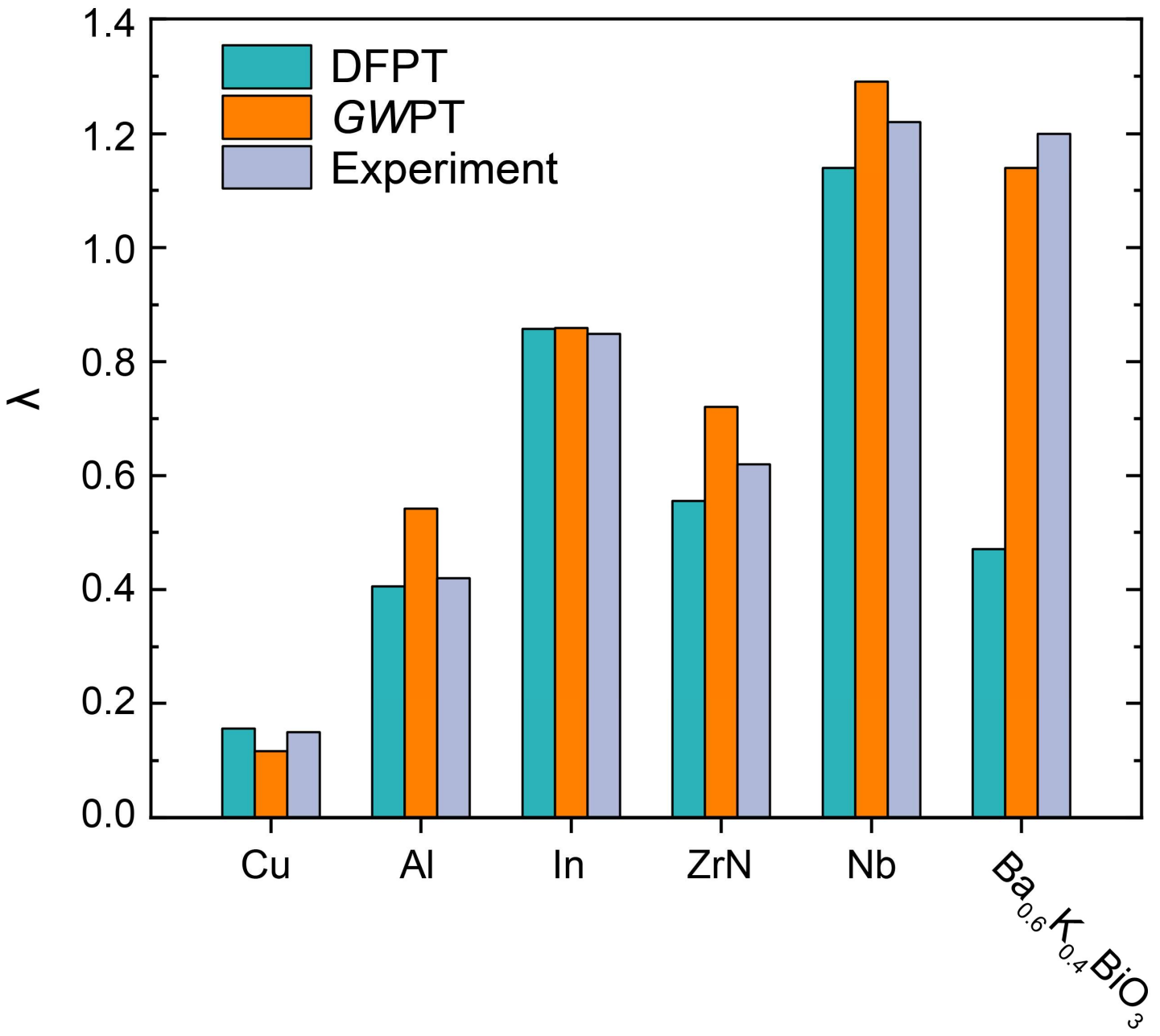

FIG. S3. DFPT- and *GW*PT-computed *e*-ph coupling constant $\lambda$ for a validation set of *sp*-band conventional metals (Cu, Al, In) and *d*-band conventional metals (ZrN, Nb), along with the oxide metal BKBO [20]. Experimental values are taken from tunneling experiments [47] for Cu [48], Al [49], In [50], ZrN [51], Nb [52], and BKBO [53], respectively. For the conventional metals studied here, DFPT and *GW*PT both agree with experiments within the experimental and numerical uncertainties. For materials with strong correlations, e.g. BKBO, DFPT fails significantly whereas *GW*PT successfully capture the many-electron effects in the measured $\lambda$ [20].